\documentclass[pre,showpacs]{revtex4}
\usepackage{epsf}
\usepackage{latexsym}

\newcommand {\vtd}[2] {{\lh\!\bay{c}#1\\#2\!\eay\rh}}

\newcommand {\beq } {\begin{equation} } \newcommand {\eeq } {\end{equation} }
\newcommand {\bit } {\begin{itemize}  } \newcommand {\eit } {\end{itemize}  }
\newcommand {\ben } {\begin{enumerate}} \newcommand {\een } {\end{enumerate}}
\newcommand {\bay } {\begin{array}    } \newcommand {\eay } {\end{array}    }

 \newcommand {\noi } {\noindent      }
\newcommand {\hsp } {\hspace          } \newcommand {\vsp } {\vspace        }
\newcommand {\hsc } {\hspace*{1cm}    }
\renewcommand {\ol} {\overline        } \newcommand {\nl  } {\newline       }

 \newcommand {\ha  } {{1\over 2}}
 \newcommand {\ov  } {\over     }
\newcommand {\e    } {\!+\!                  } \newcommand {\m   } {\!-\!     }
\newcommand {\fa   } {\forall                } \newcommand {\ev  } {\equiv    }

\newcommand {\lh  } {\left (      } \newcommand {\rh  } {\right)      }
\newcommand {\lv  } {\left[       } \newcommand {\rv  } {\right]      }
\newcommand {\lc  } {\left\{      } 
 \newcommand {\rp  } {\right.      }

\newcommand {\al} {\alpha} \newcommand {\be} {\beta} \newcommand {\la} {\lambda}
\newcommand {\cO  } {{\cal   O }}   \newcommand {\cZ  } {{\cal   Z }}
\newcommand {\cI  } {{\cal   I }}   \newcommand {\cW  } {{\cal   W }}
\newcommand {\cV  } {{\cal   V }}   \newcommand {\cE  } {{\cal   E }}
\newcommand {\cN  } {{\cal   N }}   \newcommand {\cL  } {{\cal   L }}
\newcommand {\bfn } {{\bf    n }}   \newcommand {\bfr } {{\bf    r }}
\newcommand {\bft } {{\bf    t }}   \newcommand {\bfz } {{\bf    z }}
\newcommand {\bfs } {{\bf    s }}   \newcommand {\bfx } {{\bf    x }}

\begin{document}

\title{The polynomial error probability for LDPC codes}

\author{J. van Mourik}
\affiliation{
The Neural Computing Research Group,  Aston University,
Birmingham B4 7ET, United Kingdom}

\author{Y. Kabashima}
\affiliation{
Department of Computational Intelligence and Systems Science,
Tokyo Institute of Technology, Yokohama 2268502, Japan}

\date{\today}

\begin{abstract}
We obtain exact expressions for the asymptotic behaviour of the average
probability of the block decoding error for ensembles of regular low density
parity check error correcting codes, by employing diagrammatic techniques. 
Furthermore, we show how imposing simple constraints on the code ensemble
(that can be 
practically 
implemented in linear time), 
allows one to suppress 
the error probability for codes with more than $2$ checks per bit, to an
arbitrarily low power of $N$. As such we provide a practical route to a
(sub-optimal){\em expurgated} ensemble.
\end{abstract}
\maketitle

\section{Introduction}
%
Recent research in a cross-disciplinary field between the information theory
(IT) and statistical mechanics (SM) revealed a great similarity between the low
density parity check (LDPC) error correcting codes and systems of Ising spins
(microscopic magnets) which interact with each other over random 
graphs\cite{NishimoriBook,Sourlas}. 
On the basis of this similarity, notions and methods developed 
in SM were employed to
analyse LDPC codes, which successfully clarified {\em typical} properties of
these excellent codes when the code length $N$ is 
sufficiently large\cite{MKSV,NishimoriWong,Montanari}.

In general, an LDPC code is defined by a parity check matrix $A$ which
represents dependences between codeword bits and parity checks determined under
certain constraints. This implies that the performance of LDPC codes, in
particular, the probability of the {\em block} decoding error $P_B(A)$
fluctuates depending on each realization of $A$. Therefore, the average of the
decoding error probability over a given ensemble $\ol{P_B}$ is frequently used
for characterising the performance of LDPC code ensembles.

Detailed analysis in IT literature showed that $\ol{P_B}$ 
of naively constructed
LDPC code ensembles is generally composed of two terms: the first term which
depends {\em exponentially} on $N$ represents the average performance of {\em
typical} codes, whereas the second component scales {\em polynomially} with
respect to $N$ due to a polynomially small fraction of poor codes in the
ensemble \cite{Ga,MB}.
This means that even if the noise level of the communication channel is
sufficiently low such that typical codes exhibit exponentially small decoding
error probabilities (which is implicitly assumed throughout this paper),
communication performance can still be very low exhibiting an ${\cal O}(1)$
decoding error probability with a polynomially small probability when codes are
naively generated from the ensembles.
As the typical behaviour has mainly been examined so far, the polynomial
contribution from the atypical codes has not been sufficiently discussed in the
SM approach. Although this slow decay in the error probability would not be
observed for most codes in the ensemble, examining the causes of low error
correction ability of the atypical poor codes is doubtlessly important both
theoretically, and practically for constructing more reliable ensembles.

The purpose of this paper is to answer this demand from the side of SM. More
specifically, we develop a scheme to directly assess the most dominant
contribution of the poor codes in $\ol{P_B}$ on the basis of specific
kinds of graph configuration utilising {\em diagrammatic} techniques.
This significantly simplifies the evaluation procedure of $\ol{P_B}$
compared to the existing method \cite{MB}, and can be employed 
for a wider class of expurgated ensembles. 
Moreover, it provides insights that leads to a 
{\em practical} expurgation method that is also presented in this paper.
Finally, the validity of the evaluation scheme and the efficacy of the proposed
expurgation technique are computationally confirmed. 

The paper is organised as follows:\nl
-In the next section \ref{sec:dis}, we briefly review the general scenario
 of LDPC codes and introduce basic notions which are necessary for evaluating
 the error probability in the proceeding sections. \nl 
-In section \ref{sec:rep}, we introduce the various code ensembles we will work
 with, and different representations for a code construction.\nl
-In section \ref{sec:dia}, we link the probability of having a code with low
 minimal distance to the polynomial error probability, and we calculate
 the polynomial error probability by diagrammatic techniques for various code
 ensembles. As we can explicitly link it to the occurrence of short loops, the
 distribution of occurrence of such loops is also determined.\nl
-In section \ref{sec:alg}, we present a practical linear 
time (in average) algorithm to remove short loops 
from a code construction, thus reducing the polynomial error 
probability to an arbitrarily low value. 
This is backed up by numerical simulations.\nl 
-Finally, section \ref{sec:con} is devoted to a summary. \nl
-Technical details about the diagrammatic technique can be found in appendix
 \ref{sec:appA}.\nl
-Details about the link between the minimal distance and the polynomial error
 probability for various decoding schemes are presented in appendix 
 \ref{sec:appB}.
%
\section{LDPC codes, decoding error and weight of codewords
\label{sec:dis}}
%
We here concentrate on {\em regular} $(c,d,N)$ LDPC code ensembles which involve
$N$ message bits and $L\equiv cN/d$ parity checks. Given a specific code, each
message bit is involved in $c$ parity checks, and each parity check involves $d$
message bits. In practice, this dependence is represented by a parity check
matrix $A$. An encoding scheme consists in the generation of a codeword
$\bft\in\{0,1\}^N$ from an information vector $\bfs\in\{0,1\}^K$ (with $K=N-L$)
via the linear operation $\bft=G^T\bfs$ (mod 2) where $G$ is the generator
matrix that satisfies the condition $AG^T=0$ (mod 2). The code rate is then
given by $R\ev K/N$, and measures the information redundancy of the
transmitted vector.
 
Upon transmission of the codeword $\bft$ via a noisy channel, taken here to be a
binary symmetric channel (BSC), the vector $\bfr=\bft+\bfn^0$ (mod 2) is
received, where $\bfn^0\in\{0,1\}^N$ is the true channel noise. The statistics
of the BSC are fully determined by the flip rate $p\in[0,1]$:
\begin{equation}
P(n^0_i)=(1-p)\,\delta_{n^0_i,0}+p\,\delta_{n^0_i,1}
\label{noise}
\end{equation}
Decoding is carried out by multiplying $\bfr$ by $A$ to produce the syndrome
vector $\bfz=A\bfr=A\bfn^0$ (since $AG^{T}=0$). In order to reconstruct the
original message $\bfs$, one has to obtain an estimate $\bfn$ for the true noise
$\bfn^0$. One major strategy for this is {\em maximum likelihood} (ML) decoding
and is mainly focused on in this paper. It consists in the selection of that
vector $\hat{\bfn}^{ML}$ that minimises the number of non-zero elements (weight)
$w(\bfn)=\sum_{l=1}^N n_l$ satisfying the parity check equation $\bfz=A\bfn$.
Decoding failure occurs when $\hat{\bfn}^{ML}$ differs from $\bfn^0$.
The probability of this occurring is termed as the (block) decoding error
probability $P_B(A)$, which serves as a performance measure of the code
specified by $A$ given the (ML) decoding strategy.

It is worthwhile to mention that for any true noise vector $\bfn^0$, 
$\bfn=\bfn^0+\bfx$ (mod2) where $\bfx$ is an arbitrary codeword vector for
which $A \bfx=0$ (mod 2) holds, also satisfies the parity check equation
$A\bfn=A\bfn^0=\bfz$ (mod2). The set of indices of non-zero elements of $\bfx$
is denoted by $\cI(\bfx)$.
We denote the probability that a given decoding strategy (DS) will select
$\bfn=\bfn^0+\bfx$ with $w(\bfx)=w$ rather than $\bfn^0$, as $P_{DS}(e|w)$.
The ML decoding strategy fails in correctly estimating those noise vectors
$\bfn^0$ for which less than half of $n^0_{i\in\cI(\bfx)}$ are zero, since the
weight of $\bfn^0+\bfx$(mod2) becomes smaller than $w(\bfn^0)$. To noise vectors
$\bfn^0$ for which exactly half of $n^0_{i\in\cI(\bfx)}$ are zero, we attribute
an error $1/2$, since the weight of $\bfn^0+\bfx$(mod2) is equal to
$w(\bfn^0)$, such that
\beq
P_{ML}(e|w)=\hsp*{-5mm}\sum_{i=1}^{{\rm int}((w\m1)/2)}\vtd{w}{i}p^{w-i}(1-p)^i
\lh~+_{(w~{\rm even})} {1\ov2}\vtd{w}{w\ov2}
{(p(1-p))}^{w/2}
\rh\sim\cO(p^{w\ov2})
\label{eq:PewML}
\eeq
where ${\rm int}(x)$ is the integer part of $x$ (for $P(e|w)$ for other decoding
schemes we refer to appendix \ref{sec:appB}). The minimum of $w(\bfx)$ under
the constraints $A\bfx=0$ (mod2) and $\bfx\ne0$, is commonly known as the {\em
minimal distance} of $A$ and is here denoted as ${\cal W}(A)$. It provides a
lower bound for the decoding error probability of the ML scheme as
$P_B(A)>{\cO}\lh p^{{\rm int}(\cW(A)/2)}\rh$.

Gallager \cite{Ga} showed that for $c \ge 3$ the minimal distance grows as
${\cal O}(N)$ for most codes characterised by the $(c,d)$-constraints, which
implies that the decoding error probability can decay exponentially fast with
respect to $N$ when $p$ is sufficiently low. However, he also showed that the
minimal distance and, more generally, the weights of certain codeword vectors
become ${\cal O}(1)$ for a polynomially small fraction of codes when the code
ensemble is naively constructed, which implies that the average decoding error
probability over the ensemble $\ol{P_B}$ exhibits a slow polynomial decay
with respect to $N$ being dominated by the atypical poor codes. 
As Gallager did not examine the detailed properties of the poor codes, it was
only recently that upper- and lower-bounds of $\ol{P_B}$ were evaluated
for several types of naively constructed LDPC code ensembles \cite{MB}. 
However, the obtained bounds are still not tight in the prefactors.
In addition, extending the analysis to other ensembles does not seem
straightforward as the provided technique is rather complicated.
The first purpose of this paper is to show that one can directly evaluate the
leading contribution of $\ol{P_B}$ by making a one-to-one connection
between occurrence of low weights in codeword vectors and the presence of some
dangerous finite diagram(s) (sub-graph(s)) in a graph representation of that
code.

In order to suppress the influence of the atypical poor codes, Gallager proposed
to work in an {\em expurgated} ensemble, where the codes with low minimal
distance are somehow removed. However, a practical way to obtain the expurgated
ensemble has not been provided so far. The second purpose of the current paper
is to provide a (typically) linear-time practical algorithm for the expurgation
and we numerically demonstrate its efficacy.
%
\section{code ensembles and representations of code constructions
\label{sec:rep}}
%
\subsection{Code ensembles}
%
As mentioned in the previous section, evaluating the distribution of low
weights of the poor codes in a given ensemble becomes relevant for the
current purposes. This distribution highly depends on the details of the
definition of code ensembles. We here work on the following three ensembles
defined on the basis of bipartite graphs (Fig. \ref{fig:bi}):
\begin{figure}[h]
\setlength{\unitlength}{1mm}
\begin{picture}(160,50)
\put( 50  , 5  ){\epsfysize=40\unitlength\epsfbox{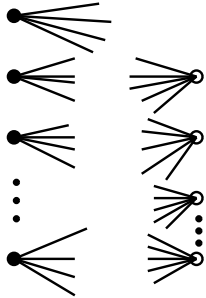}}
\end{picture}
\caption{An example of a regular bipartite graph with $(c,d)=(4,5)$. On the
left the vertices (message bits) (full circles), and on the right the edges
(parity checks) (empty circles)}
\label{fig:bi}
\end{figure}
\bit
\item {\bf Miller-Burshtein (MB) ensemble:} Put $N$ vertices (message bits) 
and $L$ edges (parity checks) on the left and right, respectively. For each
vertex and edge, we provide $c$ and $d$ arcs, respectively. In order to
associate these, the arcs originating from the left are labelled from $1$ to
$Nc(=Ld)$, and similarly done for the right. A permutation $\pi$ is then
uniformly drawn from the space of all permutations of $\{1,2,\ldots,Nc\}$.
Finally, we link the arc labelled $i$ on the left with the arc labelled $\pi_i$
on the right. This defines a code completely determining a specific dependence
between message bits and parity checks. The uniform generation of $\pi_i$
characterises the ensemble. Note that in this way multiple links between a pair
of vertex/edge are allowed.
\item {\bf No multiple links (NML) ensemble:}
Multiple links in the bipartite graph possibly reduces the effective number of
parity checks of the associated message bits, which may make the error
correction ability weaker. The second ensemble is provided by expurgating graphs
containing multiple links from the MB ensemble.
\item {\bf Minimum loop length $\ell$ (MLL-$\ell$) ensemble:}
In the bipartite graph, length $\ell$ loops are defined as irreducible closed 
paths composed of $\ell$ different vertices and $\ell$ different edges
\footnote{Length $\ell$ under this definition is usually counted as $2\ell$ in
the IT literature \cite{Loop1,McGowan}.}. 
As shown later, short loops become a cause of poor error correction ability, 
as they allow for a shorter minimal distance. Therefore, we construct code
ensembles by completely expurgating graphs containing loops of length shorter
than $\ell$ from the MB ensemble, and examine how well such expurgation
contributes to the improvement of the average error correction capability. Note
that the MB and NML ensemble correspond to the MLL-$1$ and MLL-$2$ ensemble,
respectively.
\eit
%
\subsection{Representations of code constructions}
\begin{figure}[h]
\setlength{\unitlength}{1mm}
\begin{picture}(160,40)
\put( 50  , 5  ){\epsfysize=30\unitlength\epsfbox{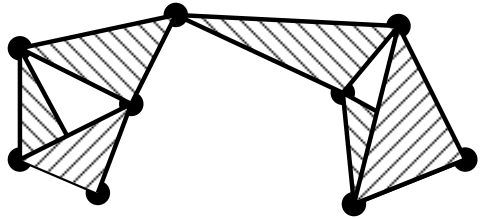}}
\end{picture}
\caption{An example of a (small) regular hyper-graph with $(c,d,N)=(2,3,9)$}
\label{fig:mono}
\end{figure}

%
Although the ensembles above are constructed on the basis of bipartite graphs,
for convenience we will also use other two representations: 
\bit
\item {\bf Monopartite (hyper)-graph representation:} 
Each message bit is represented by a vertex. 
The vertices are connected by hyper-edges (each linking $d$ vertices),
each vertex is involved $c$ times in an hyper-edge (see Fig.\ref{fig:mono}).
\item {\bf Matrix representation:}
$L$ rows, $N$ columns, where $a_{ev}$ is the number of times vertex $v$ appears
in edge $e$. For regular $(c,d,N)$ codes, the following constraints on
$\{a_{ev}\}$ apply
\beq
\lc\bay{ll}
\sum_{v=1}^N a_{ev}=d,\hsc&\fa e=1,..,L,\\
\sum_{e=1}^L a_{ev}=c,\hsc&\fa v=1,..,N.
\eay\rp
\label{reg_code}
\eeq
For clarity, we always use indices $v,w,..=1,..,N$ to indicate message bits
(or vertices), and indices $e,f,..=1,..,L$ to indicate parity checks 
(or edges).
Note that the parity check matrix of a given code is provided as $A=\{a_{ev}\}$
(mod 2). For the NML and MLL-$\ell\geq3$ ensembles, the matrix elements are
constrained to binary values as $a_{ev}=0,1$. Therefore, $\{a_{ev}\}$ itself
represents the parity check matrix $A$ in these cases, which implies that every
parity check is composed of $d$ bits and every bit is associated with $c$
checks. However, as $a_{ev}$ can take any integer from $0$ to $c$ in the MB
ensemble, it can occur that certain rows and/or columns of the parity check
matrix $A$ are composed of only zero elements, which means that corresponding
checks and/or bits do not contribute to the error correction mechanism.
Note that in the matrix notation the exclusion of $l$-loops corresponds to
the extra (redundant) constraints (additional to $a_{ve}=0,1$, and
(\ref{reg_code})):
\beq
\prod_{i=1}^l a_{v_ie_i}a_{v_{(i\e1){\rm mod}l}e_i}=0~~
\forall\{v_i,i=1..l\},~~\{e_i,i=1..l\}.
\label{nolloop}
\eeq
where $\{v/e_i,i=1..l\}$ is a group of $l$ different vertices/edges.
\eit
There is a one-to-one correspondence between the bipartite graph, and the
matrix representation of the codes. Note however, that with each monopartite
graph correspond a number of (identical up to permutation of the edges) of
bipartite graph/matrix representations.
%
\section{Diagrammatic evaluation of error probability \label{sec:dia}}
%
\subsection{Low weights and most dangerous diagrams}
%
Now, let us start to evaluate the error probability. For this, we first
investigate necessary configurations in the bipartite graph representation for
creating codeword vectors having a given weight.

Assume that a codeword vector $\bfx$ which is characterised by $A\bfx=0$ (mod 2)
has a weight $w(\bfx)=n_v$. As addition of zero elements of $\bfx$ does not
change parity checks, we can focus on only the $n_v$ non-zero elements. Then, in
order to satisfy the parity relation $A\bfx=0$ (mod 2), every edge associated
with $n_v$ vertices corresponding to these non-zero elements must receive an
even number of links from the $n_v$ vertices in the bipartite representation. 

Let us now evaluate how frequently such configurations appear in the whole
bipartite graph when a code is generated from a given ensemble. 
We refer to a sub-set of the $n_v$ vertices as $\cV_f$. Each vertex $v$ is
directly linked to a subset $\cE(v)$ of the edges. We denote $\cE(\cV_f)\ev
\bigcup_{v\in\cV_f}\cE(v)$, and $n_e\ev|\cE(\cV_f)|$. Then there are $c n_v$
{\em links} to be put between $\cV_f$ and $\cE(\cV_f)$. Note that there are
exactly $c$ links arriving at each $v\in\cV_f$, and $d$ links arriving at each
$e\in\cE(\cV_f)$. Each diagram consists in a combination $(\cV_f,\cE(\cV_f))$.
For an {\em admissible} diagram, we have the extra condition that each
$e\in\cE(\cV_f)$ receives an even number of links from $\cV_f$, such that
the bits in $\cV_f$ can be collectively flipped preserving the parity relation
$A\bfx=0$ (mod 2)).

Note that we ignore the links arriving in ${\cE (\cV_f)}$ from outside the set
$\cV_f$. For admissible diagrams, $n_e$ is limited from above by
${\rm int}({cn_v\ov2})$, where ${\rm int}(x)$ is the integer part of $x$.
A number $n_e$ of the links can be put freely, while the remaining $cn_v-n_e$
links all have to fall in the group $n_e$ (out of $L$), such that it can easily
be seen that each of those (forced) links carries a probability $\sim N^{\m1}$.

There are $\vtd{N}{n_v}\simeq N^{n_v}/n_v!$ ways of picking $n_v(\ll N)$ out of
$N$ vertices, such that each diagram consisting of $n_v$ vertices and $n_e$
edges carries a probability of occurrence proportional to $\sim
N^{n_v-(n_vc-n_e)}$.

This observation allows us to identify the ``{\em most dangerous}'' admissible
diagrams as those with a probability of occurrence with the least negative power
of $N$, i.e. that combination of $(n^*_v,n^*_e)$ that maximises $n_e-(c-1)n_v$.
The collective contributions of all other diagrams are at least a factor $1\ov
N$ smaller, and therefore negligible. From this, it immediately follows that
$n_e$ must take its maximal value which is $n_e^*={\rm int}({cn_v\ov2})$, 
while $n_v$ has to be minimised, compatible with the constraints of the code
ensemble under consideration. Hence, the probability $P_f=P_f(n^*_v)$ that a
generated graph (code) contains a {\em most dangerous} diagram including $n^*_v$
vertices, scales like
\beq
P_f(n^*_v)\sim N^{n^*_v(1-{c\ov2})}. 
\label{scale}
\eeq
with the constraint on $n^*_v$ that ${cn^*_v\ov2}$ is integer. Note that for
all ensembles we consider in this paper $n^*_v\sim\cO(1)$.
\nl
\nl
At this point, we make some important observations:
\nl
- Firstly, from eq. (\ref{scale}) it is easily seen that for $c=2$, any diagram
  containing an equal number of vertices and edges scales like $N^0$. The number
  of diagrams contributing to $\ol{P_B}$ becomes infinite, and $\ol{P_B}\sim
  \cO(1)$. It was already recognised by Gallager \cite{Ga} and \cite{MB} that
  regular $(2,d,N)$ codes have very bad decoding properties under the block
  error criterion, which is currently adopted. 
  Therefore, in what follows, we will implicitly assume that $2<c<d$.
\nl
- Secondly, as eq. (\ref{scale}) represents only the dependence on the code
  length $N$, for an accurate evaluation of the asymptotic behaviour (for
  $N\to\infty$) of the error probability, we have to calculate the prefactor.
  Nevertheless, this kind of {\em power counting} is still highly useful because
  this directly identifies the major causes of the poor performance, and allows
  us to concentrate on only a few relevant diagrams for further calculation,
  ignoring innumerable other minor factors. This is more systematic and much
  easier to apply in various ensembles than the existing method of \cite{MB}.
\nl
- Thirdly, once $P_f\simeq P_f(n^*_v)$ is accurately evaluated, the average
  block error probability $\ol{P_B}$ for any decoding scheme $DS$ and for
  sufficiently low flip rates $p$ can be evaluated as
\beq
\ol{P_{B_{DS}}}\simeq P_{B_{DS}}(e|n^*_v) P_f, 
\label{eq:P_f_P_B}
\eeq
  where e.g. for ML decoding $P_{B_{ML}}(e|n^*_v)$ is given in (\ref{eq:PewML})
  (the expressions for other decoding schemes can be found in appendix
  \ref{sec:appB}).
\nl
- The NML and MLL-$\ell$ ensembles are generated from the MB ensemble
  by expurgating specific kinds of codes, which slightly changes the distributio
  of codes such that the above argument based on free sampling of links in
  constructing graphs might not be valid. However, for $n_v\sim{\cal O}(1)$ the
  current evaluation still provides correct results for the leading contribution
  to $P_f(n_v)$, since the influence of the expurgation procedure has a
  negligible effect on the leading contributions of the most dangerous diagrams.
\nl
- Finally, we note that this analysis essentially follows the {\em weight
  enumerator} formalism \cite{MacKay,Aji}, which can be regarded as a certain
  type of the {\em annealed} approximation \cite{Anneal}. Although we have
  argued that such formalism is not capable of accurately evaluating the
  performance of typical codes that decays exponentially with respect to $N$
  \cite{vMS}, the current scheme correctly assesses the leading contribution 
  of the average error probability $\ol{P_{B}}$ of the above ensembles, which
  scales polynomially with respect to $N$ being dominated by atypically poor
  codes.
\nl
  This is because the probability of occurrence scales like $N^{\m\al}$ (with
  $\al\geq1$) for all admissible diagrams. Therefore, we can safely ignore the
  possibility that more than one such diagram occurs in the same graph (as
  illustrated in fig. \ref{fig:dia}), and to leading order for $P_f$, we can
  simply add the contributions of all most dangerous diagrams. This is not so
  for the typical case analysis (with $n^*_v\sim\cO(N)$), where exponentially
  rare codes may have an exponentially large contribution. In order to avoid
  this kind of over counting, a {\em quenched} magnetisation enumerator based
  treatment is then more suitable \cite{vMS}. Note furthermore that in the SM
  treatment of typical codes, the polynomial error probability is hidden in the
  ferro-magnetic solution (since ${w(\bfn_0)\ov N}\simeq{w(\bfn_0|n_v^*~{\rm
  flipped})\ov N}$ when $n^*_v\sim o(N)$), and is therefore easily overlooked.
%
\subsection{Evaluation of error probability for various ensembles}
%
Once the notion of most dangerous diagrams is introduced, the asymptotic
behaviour of the probability $P_f$ for sufficiently low flip rates $p$ is easily
evaluated for various ensembles.

\begin{figure}[h]
\setlength{\unitlength}{1mm}
\begin{picture}(160,50)
\put(20, 5){\epsfysize=30\unitlength\epsfbox{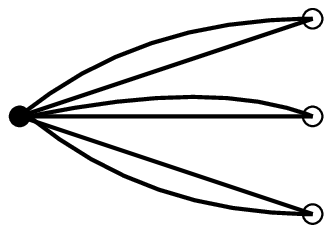}}
\put(80, 5){\epsfysize=40\unitlength\epsfbox{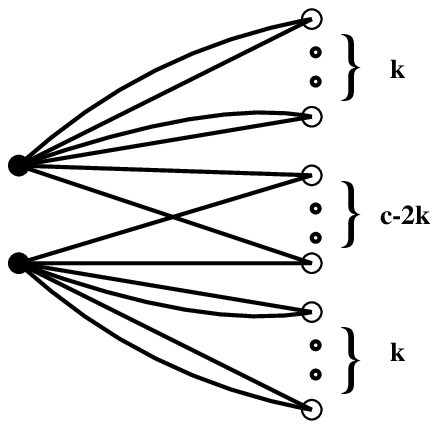}}
\end{picture}
\caption{Left: $n_v=1$, $n_e={c\ov2}$ ($c$ is even). Right: $n_v=2,~n_e=c$,
$k=0,1,..,{\rm int}({c\ov2})$}
\label{fig:lk}
\end{figure}

\begin{figure}[h]
\setlength{\unitlength}{1mm}
\begin{picture}(160,50)
\put( 10, 5){\epsfysize=40\unitlength\epsfbox{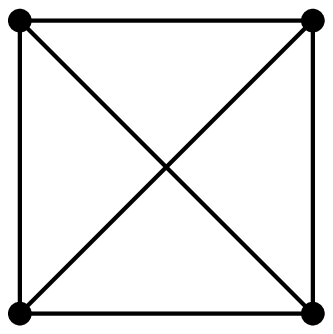}}
\put( 60, 5){\epsfysize=40\unitlength\epsfbox{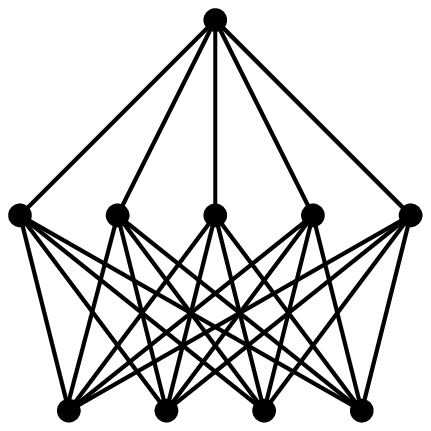}}
\put(110, 5){\epsfysize=40\unitlength\epsfbox{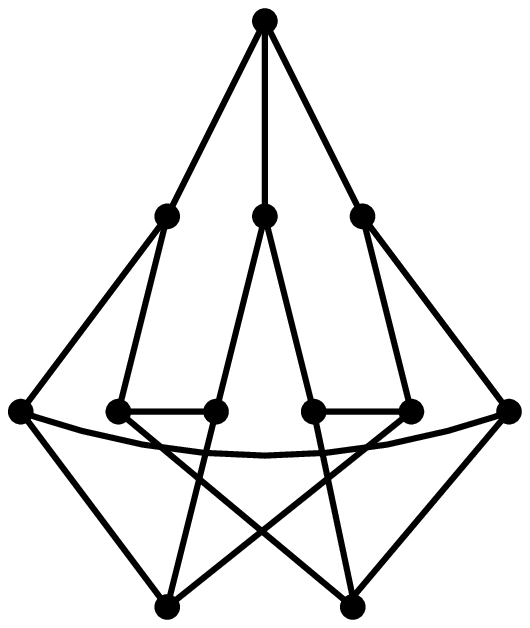}}
\end{picture}
\caption{Monopartite graph representations for some {\em most dangerous}
 diagrams.
        Left:   $c=3$, all loops of length $l<3$ are removed.
        Middle: $c=5$, all loops of length $l<4$ are removed.  
        Right:  $c=3$, all loops of length $l<5$ are removed.}
\label{fig:l345}
\end{figure}

\begin{figure}[h]
\setlength{\unitlength}{1mm}
\begin{picture}(160,70)
\put( 50  , 5  ){\epsfysize=40\unitlength\epsfbox{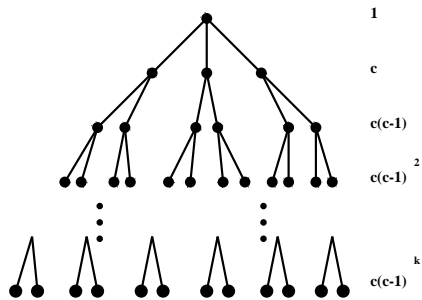}}
\end{picture}
\caption{All loops of length $l<\ell$ are removed. The minimal size of the
last generation can not be less than $c(c-1)^{{\rm int}((\ell\m1)/2)}$ without
generating loops of length $<\ell$.}
\label{fig:lg5}
\end{figure}

\begin{figure}[h]
\setlength{\unitlength}{1mm}
\begin{picture}(160,60)
\put( -5  , 5  ){\epsfysize=55\unitlength\epsfbox{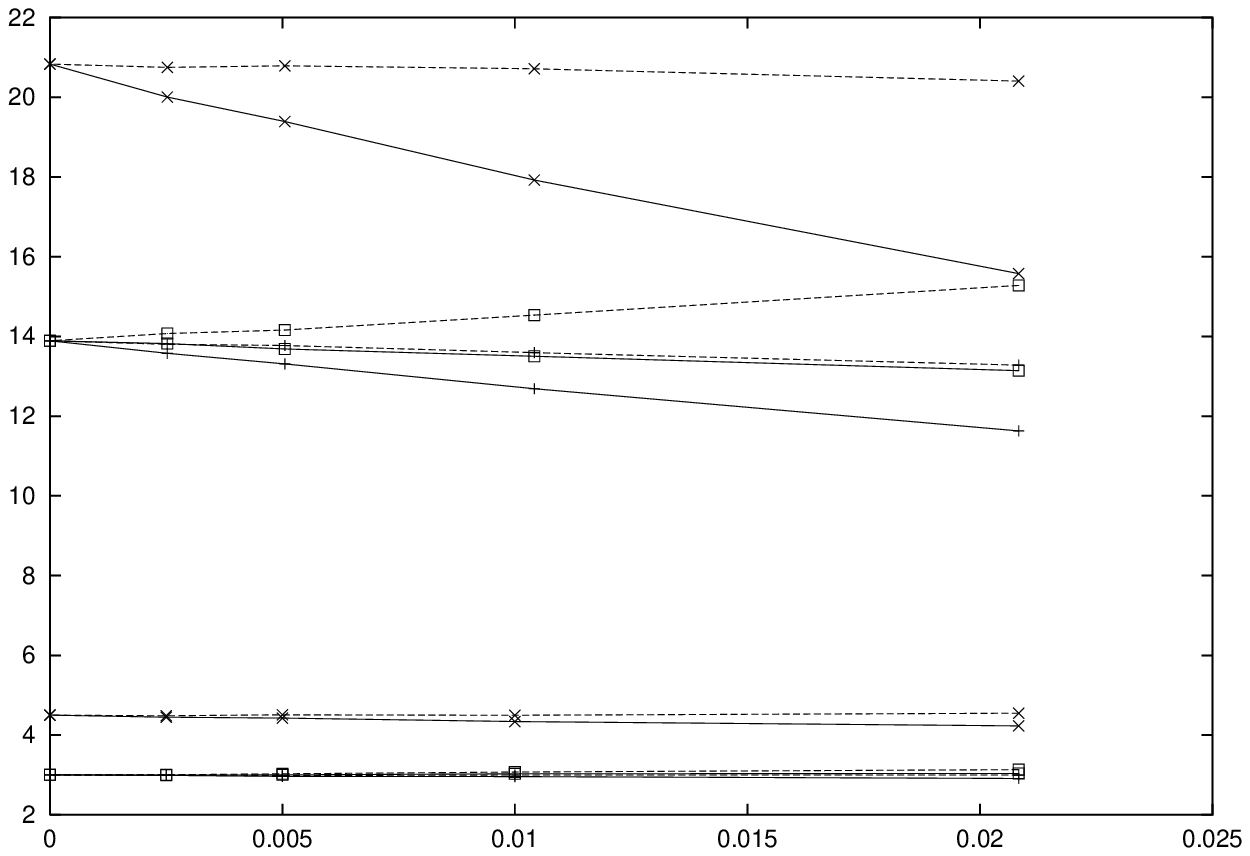}}
\put(-10  , 40 ){$f~N$}
\put( 33  , 2  ){${1\over N}$} 
\put( 80  , 5  ){\epsfysize=55\unitlength\epsfbox{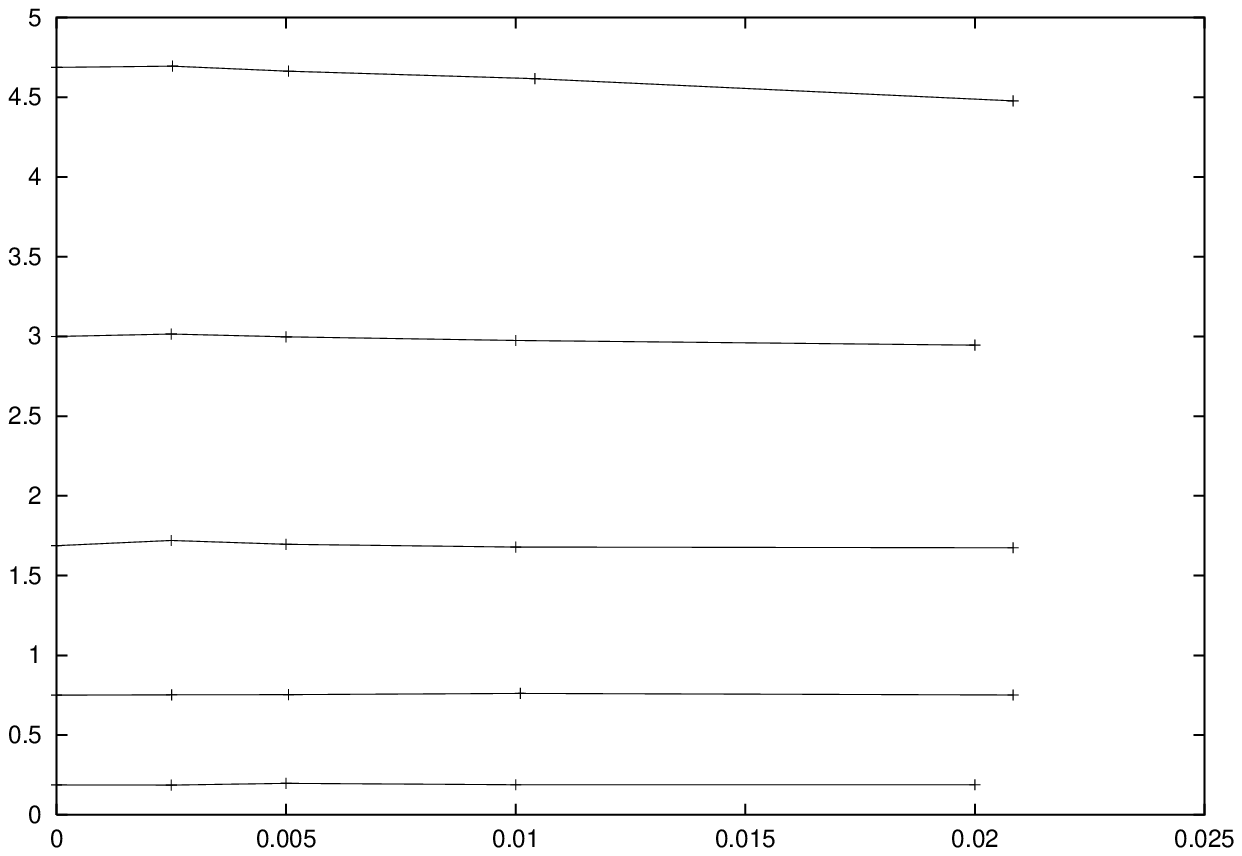}}
\put( 75  , 40 ){$f~N$}
\put(117  , 2  ){${1\over N}$} 
\end{picture}
\caption{{}Frequencies (multiplied with $N$) of occurrence of dangerous diagrams
that scale like $N^{\m1}$.
Left: $c=3$, $d=4,6$ with $k=0$ both with ($\Box$) and without ($+$) removing
1-loops, and with $k=1$ ($\times$).
Right: $c=4$, $d=2..6$.}
\label{fig:dia}
\end{figure}

$\bullet$ {\bf MB ensemble:} 
For the MB ensemble, multiple links between a pair of vertex/edge are allowed,
which forces us to make a distinction between even and odd $c$. For even $c$ the
minimal admissible $n_v$ is $1$, such that the error probability will scale like
$N^{1-{c\over2}}$, while that for odd $c$ is $2$ which provides a faster scaling
$N^{2-c}$. 
\nl
-For $c$ even, the most dangerous diagram is given in Fig.\ref{fig:lk}, and the
probability $P_f\simeq P_f(n_v=1)$ is given by (an explanation of the diagrams,
and how there multiplicity is obtained can be found in appendix A)
\beq
P_f(\ell=1,c~{\rm even})\simeq N^{1-{c\ov2}}~(1-R)^{-{c\ov2}}~
\lh{(d\m1)\ov d}\rh^{c\ov2}~{c!\ov2^{c\ov2}({c\ov2})!}~
\label{pf1e}
\eeq
Note that with ``$x\simeq y$'', we indicate that ``$x=y~(1+{o(N)\ov N})$''.
\nl
-For $c$ odd the minimal admissible $n_v$ is $2$, such that the error
probability will scale like $N^{2-c}$. The most dangerous diagrams are given in
Fig.\ref{fig:lk} (with $k=0,1,..,{\rm int}({c\ov2})$), and the probability 
$P_f\simeq P_f(n_v=2)$ is given by (for details see appendix A) 
\beq
P_f(\ell=1,c~{\rm odd})\simeq N^{2-c}~(1-R)^{-c}~\lh{(d\m1)\ov d}\rh^c~{c!\ov2}
\lv\sum_{k=0}^{{\rm int}(c/2)}{c!\ov2^{2k}k!^2(c\m2k)!}\rv
\label{pf1o}
\eeq
One can check that the values (\ref{pf1e}) and (\ref{pf1o}), which we believe to
be exact (not bounds), satisfy the bounds given in \cite{MB}.
\nl
\nl
$\bullet$ {\bf NML ensemble:} In the NML ensemble, multiple links are not
allowed. In this ensemble, even and odd $c$ can be treated on the same footing. 
In both cases, the minimal admissible $n_v=2$, such that the error probability
will scale like $N^{2-c}$. The most dangerous diagram is given in
fig.\ref{fig:lk} (note that in this case only $k=0$ is allowed). The probability
$P_f\simeq P_f(n_v=2)$ is given by (for details see appendix A) 
\beq
P_f(\ell=2)\simeq N^{2-c}~(1-R)^{-c}~\lh{(d\m1)\ov d}\rh^c~{c!\ov2}~
\label{pf2}
\eeq
\nl
\nl
$\bullet$ {\bf MLL-$3$ ensemble:} In the MLL-$3$ ensemble, neither multiple
links nor loops of length $2$ are allowed. Note that this also implies that
pairs of vertices may not appear more than once together in a parity check, such
that all parity checks are different, making each monopartite graph correspond
to exactly $L!$ bipartite graphs/matrices. In the monopartite graph any pair of
vertices is now connected by at most one (hyper-)edge.
\begin{figure}[h]
\setlength{\unitlength}{1mm}
\begin{picture}(160,50)
\put(50, 5){\epsfysize=40\unitlength\epsfbox{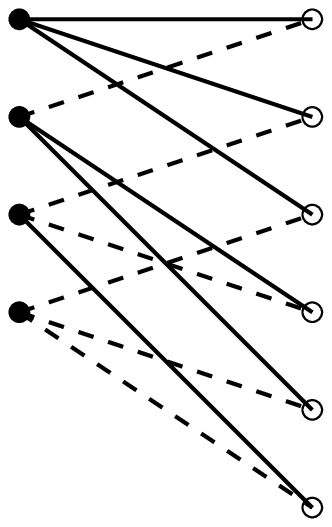}}
\end{picture}
\caption{$\ell=2$-loops removed, $n_v=c\e1$, $n_e={c(c\e1)\ov2}$.}
\label{fig:l3}
\end{figure}
One can easily convince oneself that the minimal $n_v=c+1$ (each $v$ needs $c$
other vertices to connect to), and the most dangerous diagram is given in
Fig.\ref{fig:l3}. The dominant part of $P_f\simeq P_f(c\e1)$ is given by
(for details see appendix A)
\beq
P_f(\ell=3)\simeq N^{c\e1-{c(c\e1)\ov2}}(1-R)^{-c(c\e1)\ov2}
\lh{(d-1)\ov d}\rh^{c(c\e1)\ov2}~{c!^{c\e1}\ov(c\e1)!\lh{c(c\e1)\ov2}\rh!}
\label{pf3}
\eeq
\nl
$\bullet$ {\bf MLL-$\ell$ ensemble:} In the general MLL-$\ell$ ensemble, neither
multiple links nor loops of length $l<\ell$ are allowed. In general, identifying
the most dangerous diagram(s) and especially calculating their combinatorial
prefactor becomes increasingly difficult with increasing $\ell$, but we can
still find the scaling of $P_f$ relatively easily by {\em power counting}.
To this purpose it is more convenient to use the monopartite graph
representation. In Fig.\ref{fig:l345} we observe that for $\ell=3,4$ and $5$,
the minimal $n_v$ is given by $c\e1,~2c$ and $2c+c(c\m1)=c(c\e1)$ respectively,
such that using (\ref{scale}) we obtain
\beq
P_f(\ell=3)\sim N^{(c\e1)(1-{c\ov2})}\hsc
P_f(\ell=4)\sim N^{2c(1-{c\ov2})}\hsc
P_f(\ell=5)\sim N^{c(c\e1)(1-{c\ov2})}
\label{pf345}
\eeq
For $\ell\geq6$, even finding the most dangerous diagram and thus power counting
becomes quite difficult to do by hand, but we can still easily upper bound the
power by the following observation: from fig.\ref{fig:lg5} we observe that for a
given minimal allowed loop length $\ell$ the minimum number of {\em generations}
without links between them starting from any vertex $v$ is given by
int$({\ell\m1\ov2})$. Therefore the minimal $n_v$ is lower bounded by
\beq
n_v\geq 1+c\sum_{k=0}^{{\rm int}({\ell\m1\ov2})}(c\m1)^k=
        1+c\lh{(c\m1)^{{\rm int}({\ell\e1\ov2})}-1\ov(c\m1)-1}\rh, 
\label{nvlb}
\eeq
which implies that $P_f$ can be upper bounded as
\beq
P_f(\ell)\leq\cO\lh N^{\lh 1+c\lh{(c\m1)^{{\rm int}({\ell\e1\ov2})}-1
                         \ov(c\m1)-1}\rh \rh(1-{c\ov2})}\rh.
\label{Pflb}
\eeq
\nl
In fig. \ref{fig:dia} we have plotted the frequencies of occurrence of dangerous
diagrams that scale like $N^{\m1}$. We have randomly generated $10^6$ code
realisations (for $N\simeq50,100,200,400$ i.e. the nearest integer for which
$L=cN/d$ is also integer), and have plotted both the total frequency (multiplied
by $N$) of occurrence of a diagram (dashed lines), and the frequency that a
graph contains the diagram at least once (full lines). Note that in the limit
$N\to\infty$, both coincide, illustrating the fact that we can safely ignore the
possibility that more than one such diagram occurs in the same graph.
We observe that the extrapolations ${1/N}\to0$ are all in full accordance with
the theoretical predictions.

All this clearly illustrates how the exclusion of short loops reduces $P_f$, and
thus through (\ref{eq:P_f_P_B}) the polynomial error probability probability.
Furthermore, from figs. \ref{fig:lk}-\ref{fig:lg5}, it is clear that all most 
dangerous diagrams contain short loops. Knowledge of the distribution of the
number of short loops in the various ensembles is therefore relevant for our
current purposes, and we analyse the distribution of the number of $\ell$-loops
in the next subsection.
%
\subsection{The distribution of the number of $\ell$-loops}
%
In this subsection we investigate the distribution $P_\ell(k)$ of the number $k$
of $\ell$-loops for the various code ensembles.\nl
Note that we only consider irreducible $\ell$-loops, in the sense that they
are not combinations of shorter loops (i.e. they do not visit the same vertex or
edge twice). Note that an $\ell$-{\em loop} in the monopartite graph corresponds
to a $2\ell$-{\em cycle} in the bipartite graph representation \cite{MB}.
The number of irreducible $\ell$-loops in a random regular $(c,d,N)$ graph (with
$N\to\infty$) has the following distribution:
\beq
P_\ell(k)=P(\# \ell-{\rm loops}=k)\simeq{\la_\ell^k\ov k!}~\exp(-\la_\ell),\hsc
\la_\ell\ev {(c\m1)^\ell(d\m1)^\ell\ov2\ell}
\label{pl}
\eeq
For the derivation of this result we refer to appendix A. From (\ref{pl}) we
observe that the average number of short loops increases rapidly with $c$ and
$d$. Furthermore we note the symmetry between $c$ and $d$, which reflects the
edge/vertex duality which is typical for loops.\nl
\begin{figure}[h]
\setlength{\unitlength}{1mm}
\begin{picture}(160,50)
\put( 50  , 5  ){\epsfysize=40\unitlength\epsfbox{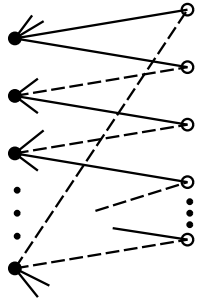}}
\end{picture}
\caption{The dominant diagram for $\ell$-loops.}
\label{fig:loop}
\end{figure}

\noi
As explained in appendix A, the constraint that no loops of length $l<\ell$ are
present in the graph, has no influence in the leading order to the diagrams for
loops of length $\geq\ell$.
\nl
\nl
We denote the number of codes in the ensemble where the minimum loop length
is $\ell$ (i.e. loops of length $l<\ell$ have been removed) by
$\cN_\ell(c,d,N)$, such that the size of the original (MB) ensemble with all
regular $(c,d,N)$ codes is denoted by $\cN_1(c,d,N)$. From (\ref{pl}) it follows
that the size of $\cN_\ell(c,d,N)$ is given by 
\beq
\cN_\ell(c,d,N)=\exp\lh-\sum_{l=1}^{\ell\m1}\la_l \rh~\cN_1(c,d,N)
        =\exp\lh-\sum_{l=1}^{\ell\m1}l{(c\m1)^l(d\m1)^l\ov2l}\rh~\cN_1(c,d,N)
\eeq
The reduction factor $\exp\lh-\sum_{l=1}^{\ell\m1}\la_l\rh$ is $\cO(1)$, for
any finite loop length $\ell$. Since the number of non-equivalent codes
in the original ensemble $\cN_1(c,d,N)\sim{(cN)!\ov(c!)^N L!}$, the final
ensemble $\cN_\ell(c,d,N)$ is still very big, but clearly smaller than
Gallager's ideally expurgated ensemble which has a reduction factor
of just ${1\ov2}$ \cite{Ga}.
\nl
Note that the presence of (short) loops in a code does not only adversely 
affect
the polynomial error probability, but also the success rate for practical
decoding algorithms such as belief progagation \cite{MacKay,McGowan}.
%
\section{Practical Linear Algorithm to the $\ell$-loop expurgated ensemble
\label{sec:alg}}
%
\noi
In this section we propose a linear time (in $N$) algorithm that generates
codes and removes loops up to arbitrary length (the combination $(c,d,N)$
permitting). We also present simulation results, which corroborate our
assumptions about the validity of the diagrammatic approach as presented in this
paper. Finally we give some practical limits and guidelines for code-ensembles
with large but finite $N$.
%
\subsubsection{Generating a random regular $(c,d,N)$ code:}
%
\noi
The algorithm to generate a random regular $(c,d,N)$ code consists in the
following steps: 
\ben
\item make a list of available vertices $A_v$ of initial length $N_{av}=cN$,
  where each vertex appears exactly $c$ times
\vsp*{-2mm}
\item for each of the $L={cN\ov d}$ parity checks, $d$ times:
 \ben
  \item  randomly pick a vertex from $A_v$,
\vsp*{-2mm}
  \item  remove it from the list
\vsp*{-2mm}
  \item  $N_{av}=N_{av}-1$.
 \een
\een
Note that in the process we keep construct the lists:
\ben
\item $EV[v][i],~v=1..N,~i=1..c$ containing the edges each vertex $v$ is
      involved in,
\vsp*{-2mm}
\item $VE[e][j],~e=1..L,~j=1..d$ containing the vertices each edge $e$ 
      involves. 
\een 
It is clear that this algorithm is linear in $N$.
%
\subsubsection{Finding loops of length $\ell$ in the code:}
%
\noi
We now describe the algorithm to detect (and store) all $\ell(\geq 2)$-loops in
the graph:
\ben
\item we consider all the vertices as a possible starting point $v_0$ of the
loop
\vsp*{-2mm}
\item given a starting point $v_0$ {\em grow} a walk of length $\ell$. Each
growing step consists in:
 \ben
  \item take $e_l     $ from $EV[v_l]$ and check conditions for valid step
\vsp*{-2mm}
  \item take $v_{l\e1}$ from $VE[e_l]$ and check conditions for valid step
\vsp*{-2mm}
  \item if all conditions are satisfied goto next step \\
        else if possible goto the next $e_l\in EV[v_l]$ or
        $v_{l\e1}\in VE[e_l]$ \\
        else go to previous step
 \een
\item finally check whether the end point of the loop $v_\ell=v_0$, if so
      store the loop i.e. $\cL_\ell=\{(v_l,e_l),l=0..\ell-1\}$
\een
The conditions for a valid step are the following:
\ben
\item
 \ben
  \item $e_l\neq e_i,~(i=0..l\m1)$~ for $\ell>2$,
\vsp*{-2mm}
  \item $e_1>    e_0$\hsc\hsc       for $\ell=2$.
 \een
\item 
 \ben
  \item $v_l>v_0,~~~~~v_l\neq v_i~(i=1..l\m1)$ for $l=1..\ell\m1,~\ell\geq2$,
\vsp*{-2mm}
  \item $v_l>v_1$\hsc\hsc\hsc~~~               for $l=\ell\m1,~~~~\ell>2$,
\vsp*{-2mm}
  \item $v_l=v_0$\hsc\hsc\hsc~~~               for $l=\ell$.
 \een
\een
Note that the conditions $v_l>v_0$ fix the starting point of the loop, while the
conditions $v_{\ell\m1}>v_1$ for $\ell>2$, and $e_1>e_0$ for $\ell=2$, fix its
orientation. This has a double advantage: it avoids over counting, and reduces
execution time by early stopping of the growing process.\nl
For 1-loops (2-cycles), for all $v=1..N$ we simply look in $EV[v]$ for double
links to the same edge, i.e. $EV[v][i]=EV[v][j]$. Imposing that $i<j$, then
avoids double counting.
\nl
Since each vertex is connected to $c$ edges, and each edge is connected to $d$
vertices, the number of operations to check whether any vertex $v$ is involved
in a loop, remains $\cO(1)$ (compared to $N\to\infty$). As we have to check
this for all vertices, the loop finding stage of the algorithm is linear in $N$.
%
\subsubsection{Removing loops of length $\ell$ from the code:}
\begin{figure}[h]
\setlength{\unitlength}{1mm}
\begin{picture}(160,60)
\put( 50  , 0  ){\epsfysize=60\unitlength\epsfbox{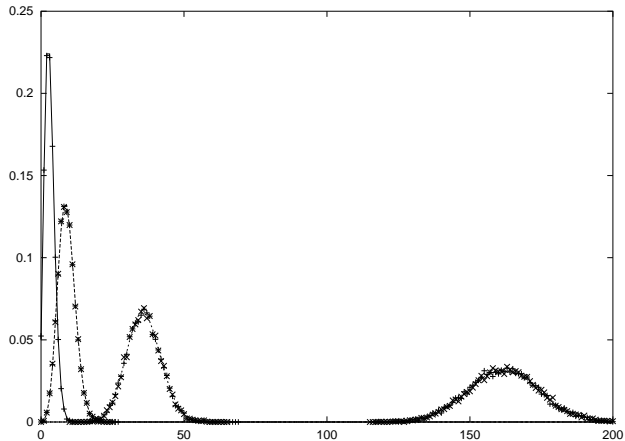}}
\end{picture}
\caption{The distribution of $\ell$-loops ($\ell=1,2,3,4$) for
$(c,d,N)=(3,4,10^4)$: lines $\to$ theory, MB ensemble $\to$ ``$+$'', MLL-$\ell$
ensemble (i.e. after removal of all smaller loops than the size plotted)
$\to$ ``$\times$'', sampled from $10^4$ random code constructions.}
\label{fig:lp34}
\end{figure}

%
We start by detecting and removing the smallest loops and than work our way
towards longer loops. Assuming that all shorter loops have been successfully
removed, and having found and listed all the loops of length $\ell$, the
procedure for removing them is very simple. For all stored $\ell$-loops
$\cL_\ell$:
\ben
\item randomly pick a vertex/edge $(v_p,e_p)$ combination from
      $\cL_\ell=\{(v_l,e_l),~l=0..\ell\m1\}$ 
\vsp*{-2mm}
\item swap it with a random other vertex $v_s$ in a random other edge $e_s$
 \footnote{After completion of this work, we found that a scheme to swap
  vertices in such a way that no new loops shorter than $\ell$ are formed based
  on the adjacency matrix $A$ was recently developed in \cite{McGowan}. However,
  it costs ${\cal O}(N^2)$ computation as computing powers of $A$ is required.
  This implies our approach, which typically runs in ${\cal O}(N)$ steps, is
  more efficient for large $N$.}.
\vsp*{-2mm}
\item for $v_p$ and $v_s$ check whether they are now involved in a $l$-loop
      with $l\leq\ell$.
 \ben
  \item if so undo the swap and goto 1.
\vsp*{-2mm}
  \item else accept the swap, the loop is removed.
 \een 
\een
This procedure of removing loops takes typically $\cO(1)$ operations.
The typical number of loops of each length $\ell$ is $\cO(1)$. For each loop we
only have to swap one vertex/edge combination to remove it, the checks that the
swap is valid take $\cO(1)$ operations, and we typically need only $\cO(1)$
swap-trials to get an acceptable swap.
\nl
Although the algorithm is linear in $N$, the number of operations needed to
detect and remove loops of length $\ell$ in a $(c,d)$ code, grows very rapidly
with $c,d$ and even exponentially with $\ell$. 
Furthermore, we note that only in the $N\to\infty$ limit all short loops can be
removed. In practice, for large but finite $N$ and given $(c,d)$, the maximum
loop length $\ell$ is clearly limited. A rough estimate for this limit is given
by
\beq
\max\lh c~{1-(c\m1)^{\ell\e1\ov2}\ov(1-(c\m1))},
        d~{1-(d\m1)^{\ell\e1\ov2}\ov(1-(d\m1))}\rh\sim N
\eeq
because $v$ (resp. $e$) is not allowed to be its own $1,2..\ell$-th nearest
neighbour (see fig. \ref{fig:lg5}). Hence, loops of logarithmic length in $N$
can not be avoided. For practical $(c,d,N)$, however, this loop length is
reached rather quickly. Therefore, we have built in the possibility for the
algorithm to stop trying to remove a given loop when after {\tt max-swap}
trials, no suitable swap has been obtained. By choosing {\tt max-swap}
sufficiently large, the maximum removable loop length is easily detected. In
practice we find that for all loop lengths that can be removed, we typically
need 1, and occasionally 2 trial swaps per loop.
\nl
In fig.\ref{fig:lp34}, we show the distribution of loops over the $(c,d,N)$
ensemble, for $(c,d)=(3,4)$, for $N=10^4$ and averaged over $10^4$ codes, up to
loops of length $\ell=4$ (corresponding to length $8$ cycles in the IT
terminology \cite{Loop1}), before and after removal of shorter loops. In
general we observe that the Poisson distribution with $\la$ given in (\ref{pl})
fits the simulations very well, for all $\ell$ not exceeding the maximal
removable loop length, while it breaks down above that.

Note that,in principle, this method also could be used to obtain Gallager's
ideally expurgated ensemble. We would start by finding and removing the most
dangerous diagrams, and then move on to the next generation of most dangerous
diagrams, and so on. However, the next most dangerous diagrams are obtained by
adding additional vertices (and necessary edges) and/or by removing edges from
the current most dangerous diagrams. One can easily convince oneself that the
number of next most dangerous diagrams soon becomes enormous. In addition for
each generation we only reduce the polynomial error probability by a factor
$N^{\m1}$. Therefore, although in principle possible, this method is not
practical, and we have opted for the removal of loops. The fact that we only
have to look for one type of diagram (i.e. loops), and the fact that we
expurgate many entire generations of next most dangerous diagrams in one go,
makes the cost of over-expurgating the ensemble a small one to pay.

\section{Summary \label{sec:con}}
%
In summary, we have developed a method to directly evaluate the asymptotic
behaviour of the average probability with respect to the block decoding error
for various types of low density parity check code ensembles using diagrammatic
techniques. The method makes it possible to accurately assess the leading
contribution with respect to the codeword length $N$ of the average error
probability which originates from a polynomially small fraction of poor codes in
the ensemble, by identifying the most dangerous admissible diagrams in a given
ensemble by a power counting scheme. The most dangerous diagrams are
combinations of specific types of multiple closed paths (loops) in the bipartite
graph representation of codes, and allow for codewords with low weights.
The contribution of a diagram to the error probability becomes larger as the
size of the diagram is smaller, which implies that one can reduce the average
error probability by excluding all codes that contain {\em any} loops shorter
than a given threshold $\ell$. We have theoretically clarified how well such a
sub-optimal expurgation scheme improves the asymptotic behaviour. We have also
provided a practical algorithm which can be carried out typically in a linear
scale of $N$ for creating such sub-optimally expurgated ensembles. The numerical
experiments utilising the provided algorithm have verified the validity of the
theoretical predictions.

The current approach is relatively easy to adapt to irregularly constructed
codes \cite{irregular,Kanter}, codes over the extended fields
\cite{Davey,Galois},  other noise channels \cite{Noise,SvMS}, and other code
constructions  such as the MN codes \cite{MN}. Work in that direction is
currently underway.

\vspace*{3mm}
\noi
{\bf Acknowledgements}
{\footnotesize
Kind hospitality at the Tokyo Institute of Technology (JvM) and support 
by Grants-in-aid, MEXT, Japan, No. 14084206 is acknowledged (YK).
}


\appendix
\section{\label{sec:appA}}
%
Diagrams are finite sub-graphs. Provided that the graph is large and provided
that the correlations between the different diagrams is not {\em too strong},
we can treat them as effectively independent to leading order in $N$, even
when they have (many) vertices and/or edges in common.
It then suffices to calculate the probability of occurrence of a single
diagram, and to count how any times such a diagram could occur in the graph, in
order to extract its overall expectation, allowing us to calculate all
quantities that depend on it. 
\nl
To illustrate this, consider the following scenario. All diagrams we consider,
consist of $n_v$ vertices and $n_e$ edges, with at least $2$ links arriving to
each of the nodes from within the diagram. Suppose now that we replace a single
node (vertex or edge), with another one not from within the diagram.
Since the probability for each link to be present is $\simeq{1\ov cN}$, there
are is at least a 4 link difference between the diagrams, thus making
the correlation between them negligible to leading order.
\nl
The rules for calculating the combinatorial pre-factor of the diagram are
easily described as follows:
\nl
Consider all possible sub-groups of $n_v$ vertices and $n_e$ edges.
Calculate the probability $P_g$ that a given group of $n_v$ vertices and $n_e$
edges forms the diagram we're interested in. Since we assume that (to leading
order in $N$) these probabilities are independent for all groups, we just have
to multiply $P_g$ with all the possible ways of picking $n_v$ vertices and
$n_e$ edges from the graph (i.e. $\vtd{N}{n_v}\vtd{L}{n_e}$).
\nl
Combined this leads to the following simple {\em recipe} for the calculation 
a the contribution of a diagram to $P_f$:
\bit
\item for each vertex from which $x$ links depart, add a factor
      $N~c!/(c-x)!$. 
\vsp*{-2mm}
\item for each edge   from which $x$ links depart, add a factor
      $L~d!/(d-x)!$. 
\vsp*{-2mm}
\item for each link, add a factor
      ${1\ov cN}$.
\vsp*{-2mm}
\item divide by the number of symmetries, i.e. the number of permutations of
      vertices, edges or links, that lead to the same diagram.
\eit 
%
\subsubsection{calculation of the diagrams in fig.\ref{fig:lk} }
%
\noi
We calculate the probability $P_f(1,{c\ov2})$ that a combination of $1$
vertex $v$, and ${c\ov2}$ edges forms the left diagram fig.\ref{fig:lk}
in the following steps:
\bit
\item 1 vertex with $c$ links ($Nc!$).
\vsp*{-2mm}
\item ${c\ov2}$ edges with 2 links ($(Ld(d\m1))^{c\ov2}$).
\vsp*{-2mm}
\item $c$ links ($({1\ov cN})^c$).
\vsp*{-2mm}
\item symmetry: ${c\ov2}$ double links ($2^{c\ov2}$).
\vsp*{-2mm}
\item symmetry: permutation of the edges ($({c\ov2})!$).
\eit
\noi
So, combined we have that
\beq
P_f(1,{c\ov2})\simeq{c!\ov2^{c\ov2}({c\ov2})!}(d(d-1)/2)^{c\ov2}
                   {NL^{c\ov2}\ov(cN)^c}~,
\label{a_pf1}
\eeq
and after some reworking we obtain (\ref{pf1e}).
\nl
\nl
We calculate the probability $P_{2,c,\kappa}$ that a combination of $2$
vertices, and $c$ edges forms the right diagram in fig.\ref{fig:lk} with
$k=\kappa$ in the following steps:
\bit
\item $2$ vertices with $c$ links ($(N c!)^2$).
\vsp*{-2mm}
\item $c$ edges    with $2$ links ($(Ld(d\m1))^c$).
\vsp*{-2mm}
\item $2c$ links ($({1\ov cN})^{2c}$).
\vsp*{-2mm}
\item symmetry: permute edges in groups of $\kappa$ ($\kappa!^2$).
\vsp*{-2mm}
\item symmetry: permute edges in group  of $c-2\kappa$ ($(c-2\kappa)!$).
\vsp*{-2mm}
\item symmetry: $2k$ double links ($2^{2k}$).
\vsp*{-2mm}
\item symmetry: simultaneously permute the vertices and the groups of $\kappa$
      edges ($2$).
\eit
\noi
So, combined we have that
\beq
P_f(2,c,k)\simeq{c!\ov2}~\lv{c!\ov 2^{2k}k!^2(c-2k)!}\rv~(d(d\m1))^c~
                {N^2L^c\ov(cN)^{2c}}~,
\label{a_pfk}
\eeq
and after some reworking we obtain (\ref{pf1o}) and (\ref{pf2}).
%
\subsubsection{calculation of the diagram in fig.\ref{fig:l3} }
%
\noi
We calculate the probability $P_f(c\e1,c(c\e1)/2)$ that a combination of $c\e1$
vertices, and $c(c\e1)/2$ edges forms diagram in fig.\ref{fig:l3} in the
following steps:
\bit
\item $c\e1$ vertices with $c$ links ($(Nc!)^{c\e1}$)
\vsp*{-2mm}
\item ${c(c\e1)\ov2}$ edges with $2$ links ($(Ld(d\m1))^{c(c\e1)\ov2}$).
\vsp*{-2mm}
\item $c(c\e1)$ links ($({1\ov cN})^{c(c\e1)}$).
\vsp*{-2mm}
\item symmetry: permute vertices ($(c\e1)!$).
\vsp*{-2mm}
\item symmetry: permute edges ($\lh{c(c\e1)\ov2}\rh!$).
\eit
\noi
So, combined we have that
\beq
P_f(c\e1,c(c\e1)/2)\simeq{c!^{c\e1}\ov(c\e1)!\lh{c(c\e1)\ov2}\rh!}~
                    (d(d\m1))^{c(c\e1)\ov2}~
                    {N^{c\e1}L^{c(c\e1)\ov2}\ov(cN)^{c(c\e1)}}~,
\label{a_pf3}
\eeq
and after some reworking we obtain (\ref{pf3}).
%
\subsubsection{distribution of the number of loops of length $\ell$}
%
\noi
The probability $P_\ell(k)$ that there are $k$ loops of length $\ell$
(i.e. including $\ell$ vertices and edges of the bipartite graph), can be
calculated from diagram fig.\ref{fig:loop}. By power counting it is easily
checked that the probability for any loop length $\ell$ to occur is $\cO(1)$.
Therefore, we adapt a slightly different strategy compared to the diagrams
above, (this also illustrates where some of the rules of our {\em recipe}
originate from)
\nl
\nl
First we calculate the probability $P_{\ell,g}$ that a given group of $\ell$
vertices (and $\ell$ edges) forms a ``true'' $\ell$-loop in the following steps:
\bit
\item We order the $\ell$ vertices into a ring (${\ell!\ov2\ell}$ ways).
\vsp*{-2mm}
\item For each pair of consecutive vertices we pick on of the edges to connect
      to both ($\ell!$ ways).
\vsp*{-2mm}
\item For each vertex choose a link to each edge it is connected to ($c(c\m1)$
      ways). 
\vsp*{-2mm}
\item For each edge choose a link to each vertex it is connected to ($d(d\m1)$
      ways). 
\vsp*{-2mm}
\item The probability that a chosen left and right link are connected is given
      by ${1\ov cN}$.
\eit
So, combined we have that
\beq
P_{\ell,g}\simeq{\ell!^2\ov 2\ell}(c(c\m1)d(d\m1))^\ell\lh{1\ov cN}\rh^{2\ell}
\eeq
There are $\vtd{N}{\ell}$ ways to pick the vertices, and $\vtd{L}{\ell}$
ways to pick the edges.\nl
We want exactly $k$ of these to form a loop, and $\vtd{N}{\ell}\vtd{L}{\ell}-k$
of these not to form an $\ell$-loop, therefore:
\beq
P_\ell(k)=\vtd{N}{\ell}\vtd{L}{\ell}P_{\ell,g}^k(1-P_{\ell,g})^{\vtd{N}{\ell}
\vtd{L}{\ell}-k}\simeq {\la_\ell^k\ov k!}\exp(-\la_\ell)
\eeq
Note that the exclusion (or not) of shorter loops, has no influence on
the leading order of $P_\ell(k)$, since the probability of having a short-cut
i.e. another edge that connects 2 vertices from within the group of $\ell$
(or vertex that connects 2 edges from within the group of $\ell$), requires
2 extra links to be present which adds a factor $\simeq{1\ov(cN)^2}$ to the
probability $P_{\ell,g}$, and is therefore negligible.
%
\section{\label{sec:appB}}
As shown, for a given code ensemble, the probability $P_f$ that a finite
group of $n_v$ bits can be collectively flipped, is completely dominated by
sub-sets of size $n^*_v$, such that $P_f\ev P_f({n^*_v})$. From this we can 
then determine the polynomial error probability $\ol{P_B}$, which depends on
the decoding scheme employed. Here, we concentrate on the BSC$(p,1-p)$ for the
following decoding schemes:
\ben
\item ML decoding \cite{Ga}: Since this decoding scheme 
selects the code word with the lowest weight, 
an error occurs when the ${n^*_v}$ collectively flipped
bits have a lower weight than the original ones. 
When the ${n^*_v}$ collectively
flipped bits have an equal weight to the original ones, we declare an error with
probability ${1\ov2}$, such that one immediately obtains (\ref{eq:PewML}).
\item MPM decoding \cite{Rujan,Sourlas_MPM,Nishimori_JPSJ}: 
This decoding scheme selects the code word that
maximizes the marginal posterior, and minimizes the bit error rate (or in a
statistical physics framework that minimizes the free energy at the Nishimori
temperature \cite{Iba}). 
\nl
Effectively this attributes a posterior probability $\exp(\be Fw(\bfn))/\cZ$ to
each codeword $\bfn$, where $\be F=\ln\lh{p\ov1\m p}\rh$, and where $\cZ\ev
\sum'_{\bfn}\exp(\be Fw(\bfn))$, with $\sum'_{\bfn}$ being the sum over all code
words. Since we assume that we are in the decodable region, we have that
$\cZ\simeq\exp(\be Fw(\bfn_0))+\exp(\be Fw(\bfn_f))$ with $\bfn_f$ being
$\bfn_0$ with $n^*_v$ bits flipped. Hence, by selecting the solution with the
maximal marginal posterior probability we obtain that
$P_{MPM}(e|n^*_v)=P_{ML}(e|n^*_v)$ as given in (\ref{eq:PewML}).
\item Typical set decoding (TS) \cite{Shannon_typical,Aji}: 
This decoding scheme randomly selects a code word from the typical set. We
declare an error when a noise different from $\bfn_0$ is selected. Hence the
error probability is given by ${n_{ts}-1\ov n_{ts}}$, where $n_{ts}$ is the
number of code words in the typical set. Since we are in the decodable region,
for $n^*_v\sim \cO(1)$, the original and the flipped code word are both (and
the only) codewords in the typical set, such that
\beq
P_{TS}(e|n^*_v)=\ha.
\label{Pe_typ}
\eeq
Note that TS decoding has an inferior performance for $\ol{P_B}$ compared to ML
and MPM decoding.
\een

\end{document}